\documentclass[10pt, conference, compsocconf]{IEEEtran}

\usepackage{graphicx}
\usepackage{algorithm}
\usepackage{algorithmic}
\usepackage{booktabs}
\usepackage[cmex10]{amsmath}

\def\call{\mathcal{L}}

\def\<{\langle}
\def\>{\rangle}

\begin{document}
\title{Social Network Intelligence Analysis to Combat\\ Street Gang Violence}
\author{\IEEEauthorblockN{Damon Paulo, Bradley Fischl, Tanya Markow, Michael Martin, Paulo Shakarian}
\IEEEauthorblockA{Network Science Center and \\
Dept. of Electrical Engineering\\
and Computer Science\\
U.S. Military Academy\\
West Point, NY 10996\\
$\{$damon.paulo, bradley.fischl, tanya.markow, michael.martin1$\}$ @usma.edu, paulo@shakarian.net}
}

\maketitle
\begin{abstract}
In this paper we introduce the Organization, Relationship, and Contact Analyzer (ORCA) that is designed to aide intelligence analysis for law enforcement operations against violent street gangs.  ORCA is designed to address several police analytical needs concerning street gangs using new techniques in social network analysis.  Specifically, it can determine ``degree of membership'' for individuals who do not admit to membership in a street gang, quickly identify sets of influential individuals (under the tipping model), and identify criminal ecosystems by decomposing gangs into sub-groups.  We describe this software and the design decisions considered in building an intelligence analysis tool created specifically for countering violent street gangs as well as provide results based on conducting analysis on real-world police data provided by a major American metropolitan police department who is partnering with us and currently deploying this system for real-world use.
\end{abstract}

\IEEEpeerreviewmaketitle

\section{Introduction}

Violent street gangs are a major cause of criminal activity in the United States~\cite{bertetto12}.  In this paper, we present a new piece of software, Organizational, Relationship, and Contact Analyzer (ORCA) that is designed from the ground-up to apply new techniques in social network analysis and mining to support law enforcement.  In particular, we look to enable improved intelligence analysis on criminal street gangs.  The software combines techniques from logic programming,~\cite{mancalogAAMAS13} viral marketing,~\cite{kleinberg,shak12} and community detection~\cite{newman04,blondel08} in a usable application custom-tailored for law enforcement intelligence support. This work is inspired by recent work in law enforcement that recognizes similarities between gang members and insurgents and identifies adaptations that can be made from current counter-insurgency (COIN) strategy to counter gang violence.~\cite{bertetto12,goode12}  Due to the striking similarities between gang-violence and COIN, the authors from the U.S. Military Academy (West Point) responded to requests from a major metropolitan police department to transition recent social network mining software.  As a result, several West Point cadets were able to not only conduct research, but also gain a better understanding of a COIN environment.

The main contribution of this paper is the ORCA software which is the first software, to our knowledge, that combine the aforementioned techniques into a single piece of software designed for law-enforcement intelligence analysis.  The paper is organized as follows.  Section~\ref{sysSec} describes ORCA and its various components while Section~\ref{evalSec} provides the results of our preliminary evaluation.  Finally, related work is discussed in Section~\ref{rwSec}.

\section{System Design and Implementation}
\label{sysSec}

The police department we worked with to develop ORCA described several issues concerning the intelligence analysis of street gangs.  They desired a software system that could accomplish the following tasks.
\begin{enumerate}
\item \textit{Ability to ingest police arrest data and visualize network representations of such data.} The police data in question primarily consists of arrest reports which include the individual's personal information as well as claimed gang membership (if disclosed).  This data also includes relationships among individuals arrested together.
\item \textit{Ability to determine degree of group membership.} While many gang members will disclose their gang affiliation, some will not - likely fearing legal consequences.  Hence, to better allocate police efforts and intelligence gathering resources, it is important to assign these unaffiliated members to a gang (with a degree of confidence).
\item \textit{Ability to identify sets of influential members.}  Though criminal street gangs are decentralized,  it is suspected that there are groups of individuals that can exert influence throughout a given gang - encouraging criminal activity that is more violent and risky than the norm.  Identifying groups of individuals in this position of influence would provide law enforcement professionals insight into the ability of these organizations to easily adopt such behavior.
\item \textit{Ability to map the ``ecosystem'' of a given gang.}  Criminal street gangs tend to be highly modular organizations, with identifiable sub-groups.  In particular, ``corner crews'' - groups of individuals in a gang who conduct illicit drug transactions on the same street corner - will form highly connected clusters in a social network representation of a street gang.  Further, understanding the relationships among these sub-organizations - both within a given gang and between different gangs provides substantial insight into inter/intra gang dynamics as well as in identifying individuals that connect different organizations.
\end{enumerate}

Note that all of the above described functions were included in ORCA in a direct response to the needs of the law enforcement professionals that we have met.  These individuals are directly from our target user population to whom we are transitioning the software to in mid-2013.
 
\begin{figure}[t!]
\centering
\includegraphics[width=0.50\textwidth]{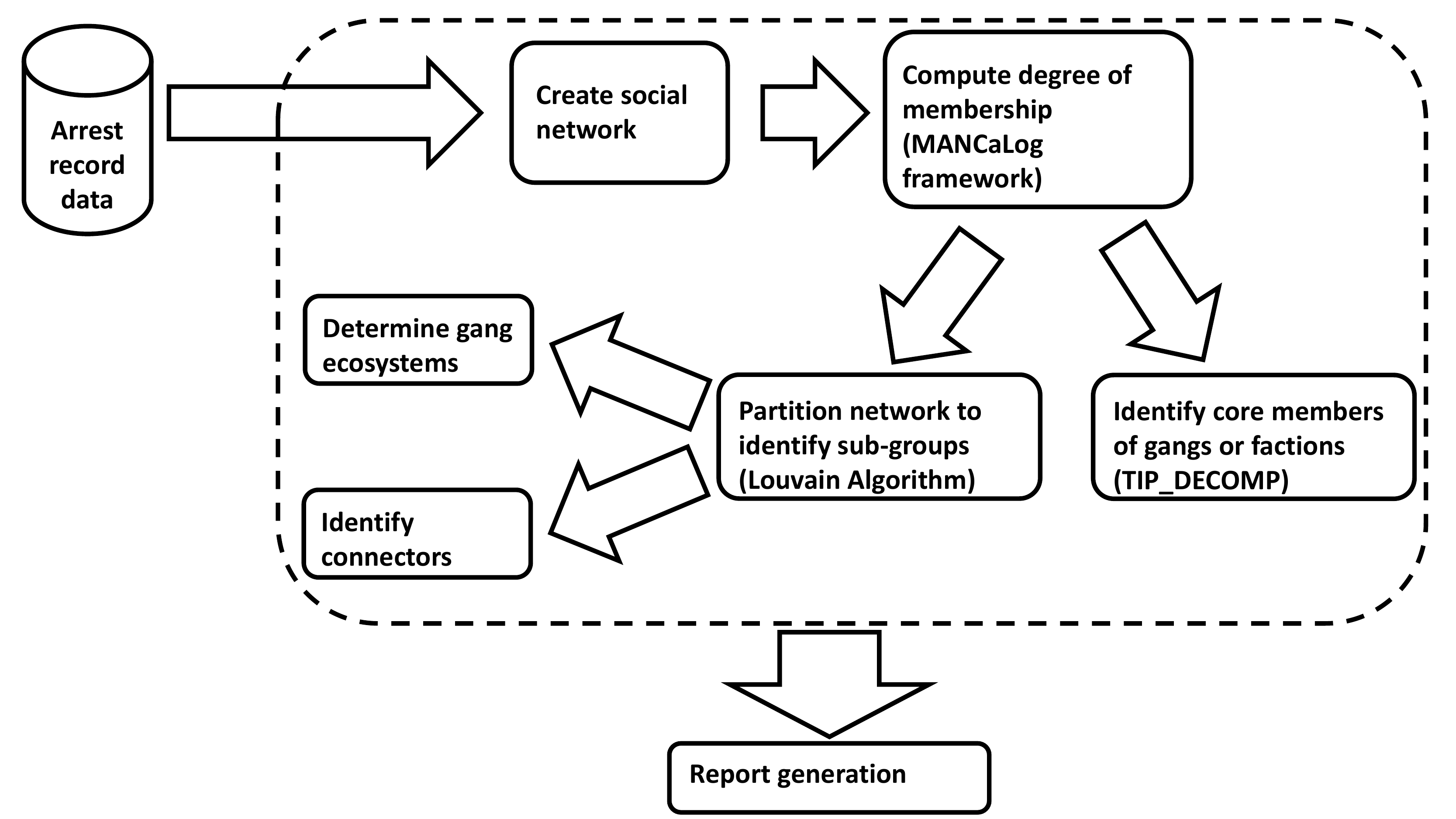}
\caption{Overview of components and functions of ORCA.}
\label{fcFig}
\end{figure}

ORCA was written in Python 2.7.3 on top of the NetworkX 1.7 library~\footnote{http://networkx.github.io/documentation/latest/index.html}.  In Figure~\ref{fcFig} we show our overall scheme for the design of ORCA to address the above challenges.  The key components we leveraged in building the system included the MANCaLog framework which we used to determine degree of membership, TIP\_DECOMP was leveraged to find sets of influential members, and the Louvain algorithm which we built on to map the ecosystems of the various gangs.  We describe how we designed each major part throughout the remainder of this section.

Police data was ingested into the system in the form of spreadsheet data that was derived from SQL queries from a police database.  For purposes of development and experimental evaluation, all personally-identifying information was anonymized.  From this data, we utilized the ``individual record'' number of each person arrested and created a social network between two individuals that were arrested together (co-arrestees).  To handle this network data structure, we utilized the Python library NetworkX mentioned earlier.  The police also required visualizations of these social networks (which are shown later).  These visualizations were also included in the context of intelligence reports automatically generated by the software (which are based on the subsequently described components).

\subsection{Determining Degree of Membership}

Analytically associating arrested individuals with a criminal gang is an important piece of intelligence to law enforcement officials.  Even though analysis itself is not direct evidence against an individual, it may be used as a starting point to gather further information.  Individuals who do not admit to being in a gang may still have associates that \textit{do} admit to membership of a certain gang.  Hence, we looked to contribute to the solution to this problem by assigning such individuals a \textit{degree of membership} - a real number in the interval $[0,1]$ that represents the confidence that the individuals is in a given gang.  A value of $0$ may be interpreted as having no information on the individual's affiliation with a given gang while a value of $1$ would be interpreted as the individual as having admitted to being in the gang.  To assign these degree of membership values, we utilize MANCaLog, a logic-programming based framework~\cite{mancalogAAMAS13}.  This framework considers a network represented as a graph $G=(V,E)$ where $V$ is a set of vertices and $E$ is a set of (undirected) relationships.  For a given $i\in V$, $\eta_i = \{j | (i,j) \in E \}$ and $d_i = |\eta_i|$.  The vertices can be assigned with a label from the set $\call$.  Each label assigned to a vertex is given a value (in the interval $[0,1]$ which associates that label with the vertex with a degree of confidence.  Hence, we use of network representation of the co-arrestees as the graph and the set of different gangs are the labels.  Individuals who are known to be in a certain gang are assigned a confidence value of $1$ for that gang and $0$ for the others.  Through MANCaLog, we can then assign a degree of confidence to the remaining nodes derived from rules of the following form:

\begin{equation}
\mathit{grp}_1 \leftarrow \bigvee_i\neg \<\mathit{grp}_i,1\>, (\mathit{grp}_1)_\mathit{ifl}
\end{equation}

Intuitively, the above rule says that a node will be assigned to group 1 if it has not already been assigned to another group (with a confidence of $1$) and has a certain number of neighbors who are also in group 1.  A set of such rules is referred to as a \textit{MANCaLog program} and denoted $P$.  A program may also have facts of the form $(\<\mathit{grp}_i,x\>,v)$ which means that vertex $v$ has a degree of membership in group $i$ with a confidence of $x$.

In the above rule, the confidence value assigned to a node being in group 1 is based on the influence function ($\mathit{ifl}$) which maps natural numbers to reals in the interval $[0,1]$.  For instance, an influence function may look something like the following:

\begin{equation}
\textit{ifl}(x) = \begin{cases} 0.0  &\textit{if } x = 0  \\ 0.1  &\textit{if } 1 \leq x  \leq 3  \\ 0.5 & \textit{if } x \geq 4 \end{cases}
\end{equation}

In \cite{mancalogAAMAS13}, the authors showed that such confidence values can be assigned to all vertices in the network for all labels in polynomial time.  However, a key issue is to derive the influence function for the rules.  We devised a simple strategy for learning the influence function for each label (in this case group) and have included it as algorithm \textsf{IFL\_LEARN}.

\algsetup{indent=1em}
	\begin{algorithm}[t!]
		\caption{ \textsf{IFL\_LEARN}}
\label{alg:dmLearn}
		\begin{algorithmic}[1]
\small
		\REQUIRE Group label $g$, program $P$
		\ENSURE
		\medskip

		\STATE{Let $R$ be an array indexed from $1$ to $d_*$ (the maximum degree of the network).}
		\FOR{$i=1,\ldots d_*$}
			\STATE{Set $pos = 0$, $tot=0$}
			\FOR{$v \in V$}
				\STATE{Set $X_v = \{ v' | (v',v) \in E \wedge (\<g,1\>,v') \in P\}$}
				\IF{$|X_v| \geq i$}
					\STATE{$tot=tot+1$}
					\IF{$(\<g,1\>,v) \in P$}
						\STATE{$pos = pos+1$}
					\ENDIF
				\ENDIF
				\ENDFOR
			\STATE{If $i=0$, $R[i]=pos/tot-1.96*SER(pos,tot)$, else $R[i]=\max(R[i-1],pos/(pos+neg)-1.96*SER(pos,tot))$ where $SER$ is the standard error on the fraction of positive neighbors.}
		\ENDFOR
		\RETURN $R$
		\end{algorithmic}
\end{algorithm}

\noindent

Algorithm~\textsf{IFL\_LEARN} considers one group label ($g$) and finds the fraction of neighbors.  We view each node as getting ``signals'' from its neighbors who are in that group.  For each potential number of signals  (which is between one and the maximum in-degree of the network) Algorithm~\ref{alg:dmLearn} below computes the fraction of nodes with at least that number of signals who are in the group.  It then computes the $95\%$ lower bound confidence interval based on standard error.  This is used as the lower bound of the degree of membership as returned by the function.  Then, based on an assumption of monotonicity (based on the results of \cite{centola10}), the algorithm sets the lower bound to be the maximum between the previous lower bound and the current.

The algorithm returns a matrix $R$, and the resulting influence function is $\mathit{ifl}(x) = R[x]$.  This algorithm is a simple approach to learning rules from available data and no novelty is claimed; more complex and general approaches to rule learning are outside the scope of this paper and are the topic of ongoing work.

\subsection{Identifying Seed Sets}

According to the law enforcement professionals we met, the idea of influence is critical to understanding the behavior of violent street gangs.  Of particular concern is the influence of radicalizing gang members - charismatic individuals that not only participate in abnormally risky and violent behavior, but have the ability to encourage others to do the same.  In order to identify sets of individuals who conduct such behavior, we consider the idea of influence spread through the ``tipping model''~\cite{Gran78}.  In such a model, we again consider a population structured as a social network ($G=(V,E)$) where each vertex $i \in V$ is adjacent to $d_i$ edges.  We model a social contagion that spreads through the network as follows: consider unaffected node $i$.  If at least $\lceil d_i/2 \rceil$ of $i$'s neighbors have the contagion, then $i$ has the contagion -- otherwise it does not.  A key question regarding this model is to identify a \textit{seed set} - a set of vertices in the network that, if initially infected, will lead to the entire population receiving the contagion.  Identifying a seed set provides the analyst a set of individuals that, when taken together, are very influential to the overall network.  Further, if the seed set is of minimal size, then the size of the set can be used as a proxy to measure how easily the network can be influenced.  Unfortunately, a simple reduction from set cover shows that finding a seed set of minimal size is an NP-hard problem.~\cite{kleinberg}  However, we have designed a fast heuristic algorithm that finds seed sets of very small size in practice.~\cite{shak12}  This algorithm is based on the idea of shell decomposition often cited in physics literature~\cite{Seidman83,ShaiCarmi07032007,InfluentialSpreaders_2010,baxter11} but modified to ensure that the resulting set will lead to all nodes being infected.  The algorithm, \textsf{TIP\_DECOMP} is presented in this section.

\algsetup{indent=1em}
	\begin{algorithm}[h!]
		\caption{ \textsf{TIP\_DECOMP}}
		\begin{algorithmic}[1]

		\REQUIRE Threshold function, $ \theta $ and directed social network $G=(V,E)$
		\ENSURE $ V' $
		\medskip

		\STATE{ For each vertex $ i,\ dist_i = \lfloor d_i/2 \rfloor $}.
		\STATE{ FLAG = TRUE}.
		\WHILE{ FLAG }
			\STATE {Let $ i $ be the element of $ V $ where $ dist_i $ is minimal}.
			\IF { $ dist_i = \infty $}
				\STATE{ FLAG = FALSE}.
			\ELSE
				\STATE{ Remove $ i $ from $G$ and for each $ j $ in $ \eta_i $, if $dist_j > 0$, set
				$  dist_j = dist_j-1 $.  Otherwise set $dist_j = \infty $}.
			\ENDIF
		\ENDWHILE
		\RETURN{ All nodes left in $ G $}.
	\end{algorithmic}
\end{algorithm}

Intuitively, the algorithm proceeds as follows (Figure 1).  Given network $G=(V,E)$,at each iteration, pick the node for which $\lfloor d_i/2 \rfloor$ is the least but positive (or $0$) and remove it.  Once there are no nodes for which this quantity is positive (or $0$), the algorithm outputs the remaining nodes in the network.  

\begin{figure}
    \begin{center}
        \includegraphics[width=1\linewidth]{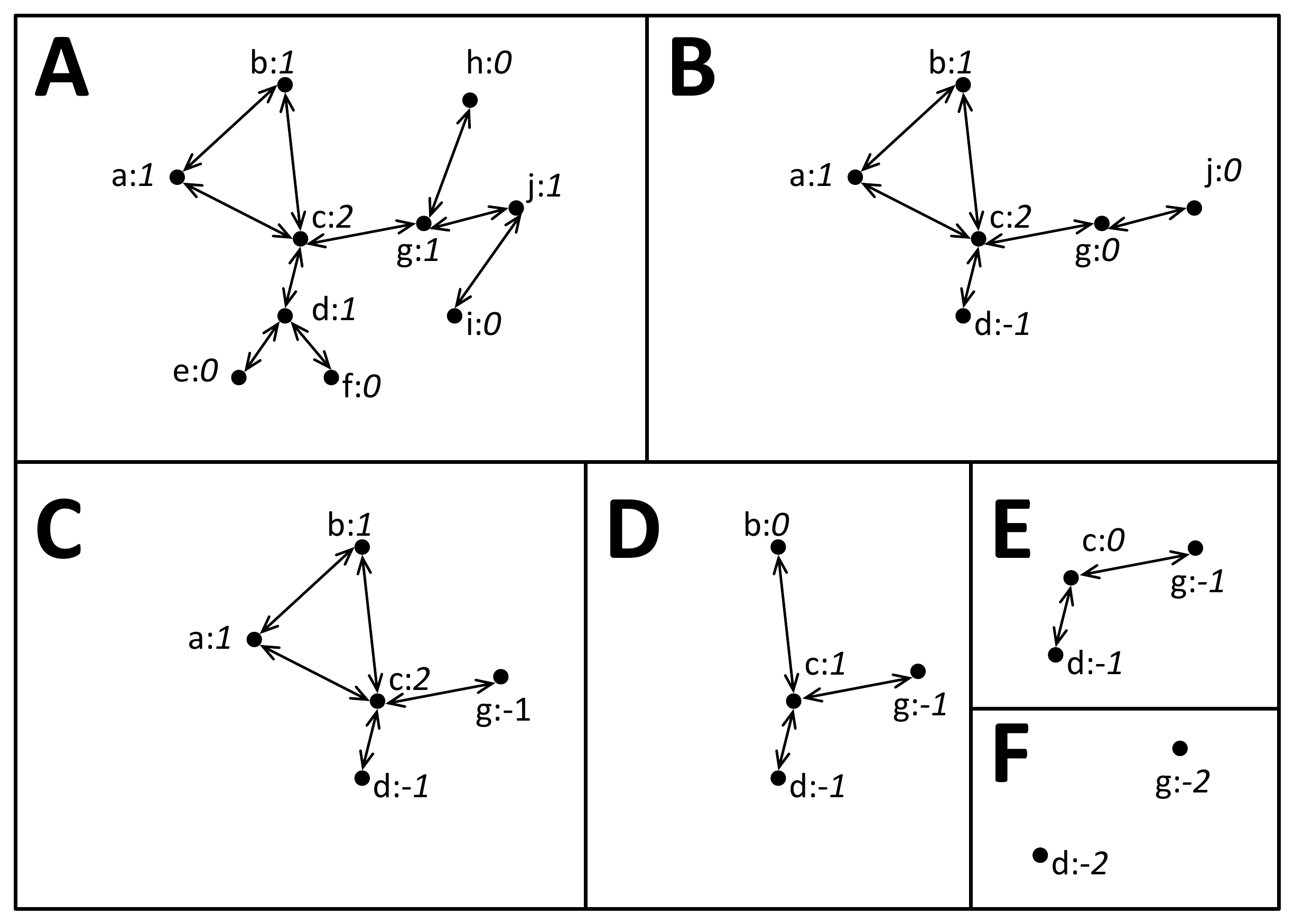}
    \end{center}
\caption{Example of \textsf{TIP\_DECOMP} for a simple network depicted in box \textbf{A}.  Next to each node label (lower-case letter) is the value for $\lfloor d_i/2 \rfloor$.  In the first four iterations, nodes e, f, h, and i are removed resulting in the network in box \textbf{B}.  This is followed by the removal of node j resulting in the network in box \textbf{C}.  In the next two iterations, nodes a and b are removed (boxes \textbf{D}-\textbf{E} respectively).  Finally, node c is removed (box \textbf{F}).  The nodes of the final network, consisting of d and g, have negative values for $\lfloor d_i/2 \rfloor$ and become the output of the algorithm.}
\end{figure}

In addition to providing the seed set for each street gang, ORCA also provides information about influential individuals.  The \textit{shell number} for each vertex based on the process of \textit{shell decomposition}~\cite{Seidman83} has been previously shown to correlate with the vertex's influence based on the networked version of various epidemic models.~\cite{InfluentialSpreaders_2010}

\subsection{Identifying Ecosystems}
Another important capability for ORCA was to decompose street gangs into component organizations.  In some cases, street gangs will not be monolithic organizations, but rather confederations of various factions - many of which may be ill identified by authorities.  Further, to maintain a gang's lines of funding, sub-organizations known as ``corner crews'' operate as an informal unit to conduct narcotics sales within certain parts of a given gang's territory.

A common method to identify communities in a social network is to partition it in a way to maximize a quantity known as modularity.~\cite{newman04}  We shall use the notation $C= \{c_1,\ldots,c_q\}$ to denote a partition over set $V$ where each $c_i \in C$ is a subset of $V$, for any $c_i,c_j \in C$, $c_i \cap c_j = \emptyset$ and $\bigcup_i c_i = V$.  For a given partition, $C$, the modularity $M(C)$ is a number in $[-1,1]$ .  The modularity of a network partition can be used to measure the quality of its community structure. Originally introduced by Newman and Girvan~\cite{newman04}, this metric measures the density of edges within partitions compared to the density of edges between partitions.  A formal definition of this modularity for an undirected network is defined as follows.

Given partition $C= \{c_1,\ldots,c_q\}$, \textbf{modularity}, 
\[
M(C) = \frac 1 {2m} \sum_{c \in C}\sum_{i,j \in c}w_{ij}-P_{ij}
\]
where $P_{ij}=\frac{k_i k_j}{2m}$.

The modularity of an optimal network partition can be used to measure the quality of its community structure.  Though modularity-maximization is NP-hard, the approximation algorithm of Blondel et al.~\cite{blondel08} (a.k.a. the ``Louvain algorithm'') has been shown to produce near-optimal partitions.\footnote{Louvain modularity was computed using the implementation available from CRANS at  http://perso.crans.org/aynaud/communities/.}  The modularity associated with this algorithm is often called the ``Louvain modularity'' and the associated partition is a ``Louvain partition.''

ORCA not only finds the Louvain partition, but also explores the relationships among the sub-groups within and between gangs.  For a given gang, ORCA generates a graph showing the relationship among all sub-groups in that gang and the neighboring sub-groups from different gangs.  We call this the ``ecosystem'' of a given gang.  The ecosystem is of particular importance to law-enforcement officers for many reasons.  One, in particular, is the issue of gang retaliation.  Consider the following: in the aftermath of a violent incident initiated by group A against group B, police enforcement will increase patrols in the territory controlled by group B.  As a result, group B will rely on an allied organization (group C) to conduct retaliatory action against group A.  Identifying the ecosystem of a given gang helps provide insight into organizations likely to conduct such activities.

Further, it identifies individuals who have connections to various other sub-groups.  The intuition behind these individuals is that they connect various organizations together.  We refer to these individuals as ``connectors.'' Connectors are also important to law-enforcement personnel as they are individuals who often connect geographic disparate sub-groups of a larger gang organization.  Another use for connectors is as a liaison between gangs that cooperate.  For instance, suppose group 1 wants to sell drugs in group 2's territory.  An agreement is reached between the two groups that group 1 can conduct these sales provided it pays a tax to group 2.   Hence, a liaison between the two groups may be a facilitator in such an arrangement.
\section{Evaluation}
\label{evalSec}

We evaluated ORCA on a police dataset of $5418$ arrests from a single police district over a three period of time.  There were $11,421$ relationships among the arrests.  From this data, ORCA assembled a social network consisting of $1468$ individuals (who were members in one of $18$ gangs) and $1913$ relationships.  ORCA was able to complete this assembly in addition to all analysis (determining degree of membership, finding seed sets, and developing ecosystems) took $34.3$ seconds on a commodity laptop (Windows 8, B960 2.2 GHz processor with 4 GB RAM).

\subsection{Degree of Membership}

\begin{figure}[t!]
\centering
\includegraphics[width=0.4\textwidth]{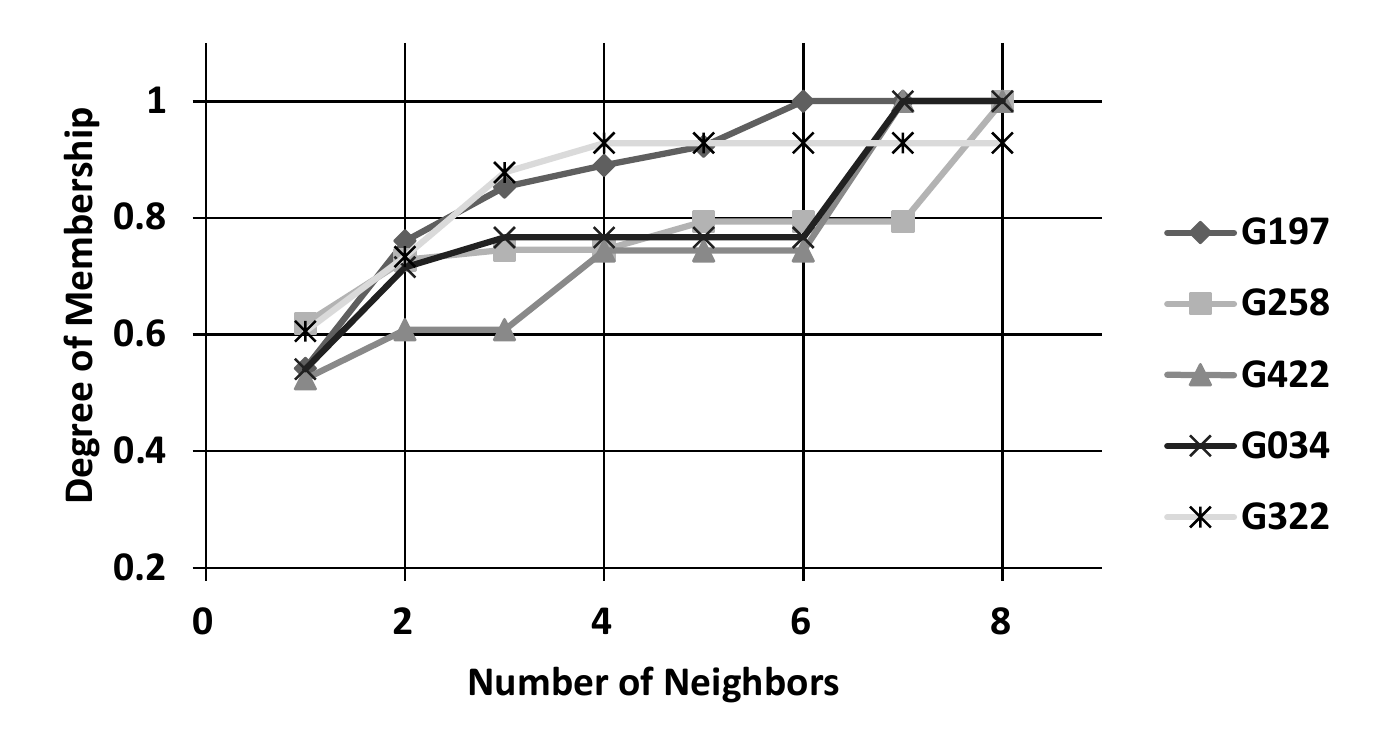}
\caption{Influence functions for five of the gangs examined by ORCA.}
\label{degFig}
\end{figure}

First, we examined the performance of the system in identifying degree of membership.  As stated earlier, this was determined using our MANCaLog framework that operated in two steps: learning the influence functions and then applying the MANCaLog inference engine to compute degree of membership. Figure~\ref{degFig} shows a plot of the influence functions derived for five of the gangs.  Specifically, it shows the number of neighboring individuals in the given gang on the x-axis compared to the computed degree of membership for an individual having that number of gang contacts on the y-axis.  We note that our degree of membership computation provides a similar result to \cite{centola10} which studies the related problem of social contagion.

\begin{figure}[t!]
\centering
\includegraphics[width=0.4\textwidth]{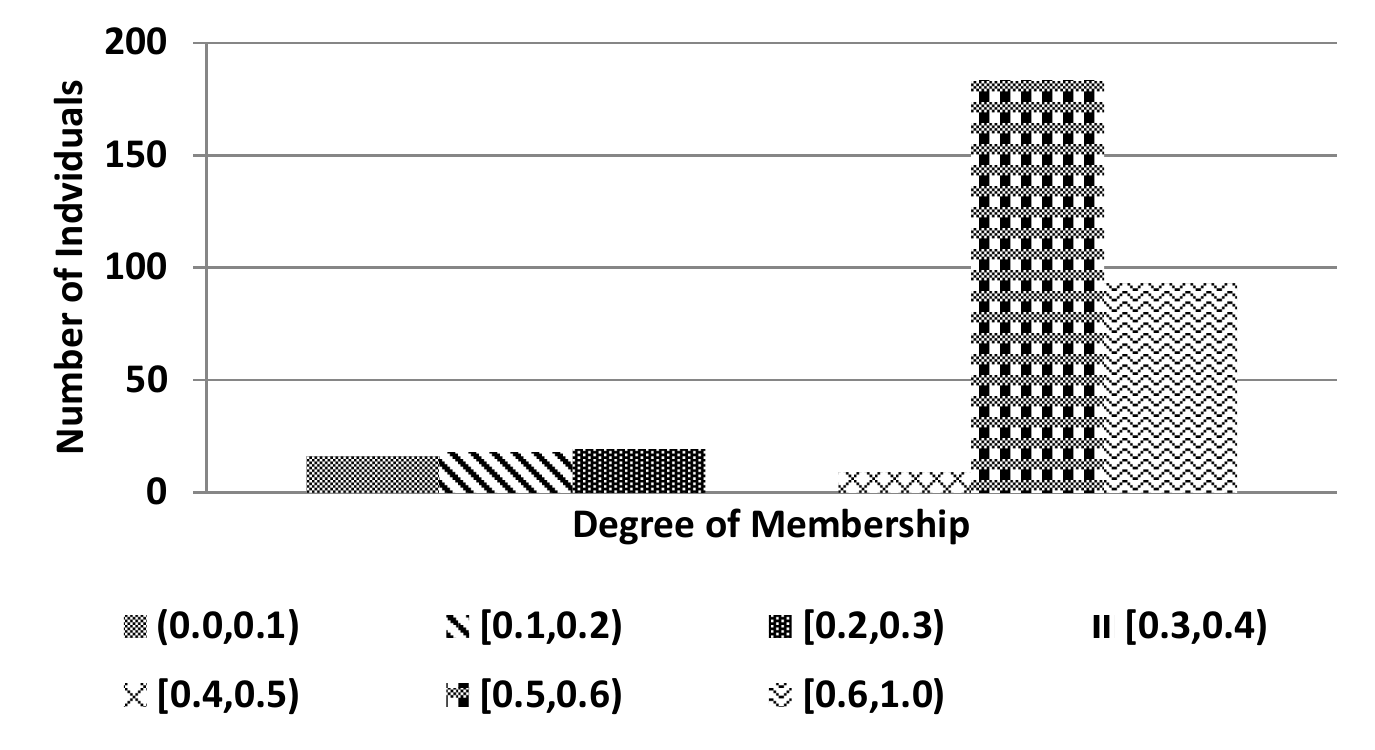}
\caption{Histogram showing the number of individuals assigned a degree of membership within a certain range.}
\label{histoFig}
\end{figure}

In the second step required to determine the degree of membership, the MANCaLog inference engine created logic programs that utilized the influence functions to determine degree of membership for individuals who did not admit to being in a gang.  All of the $180$ unadmitted individuals connected to the derived social network (individuals who did not admit to being in a gang but were arrested with at least one other individual in the district) were assigned a non-zero degree of membership based on learned rules.  The majority of these individuals could be assigned a degree of membership greater than $0.5$ for at least one gang (see Figure~\ref{histoFig}).  We note that many of these individuals were assigned a degree of membership to multiple gangs.  This may primarily be due to the fact that the included arrest reports spanned a three-year time-frame - which our police counterparts state is a relatively long period of time in street gang culture.  In such a time, gang members often change allegiances and/or form new gangs.  Exploring a longitudinal study where time is explicitly considered is an important direction for future work.  The MANCaLog framework allows for temporal reasoning as well.

\subsection{Seed Set Identification}

\begin{figure}[t!]
\centering
\includegraphics[width=0.4\textwidth]{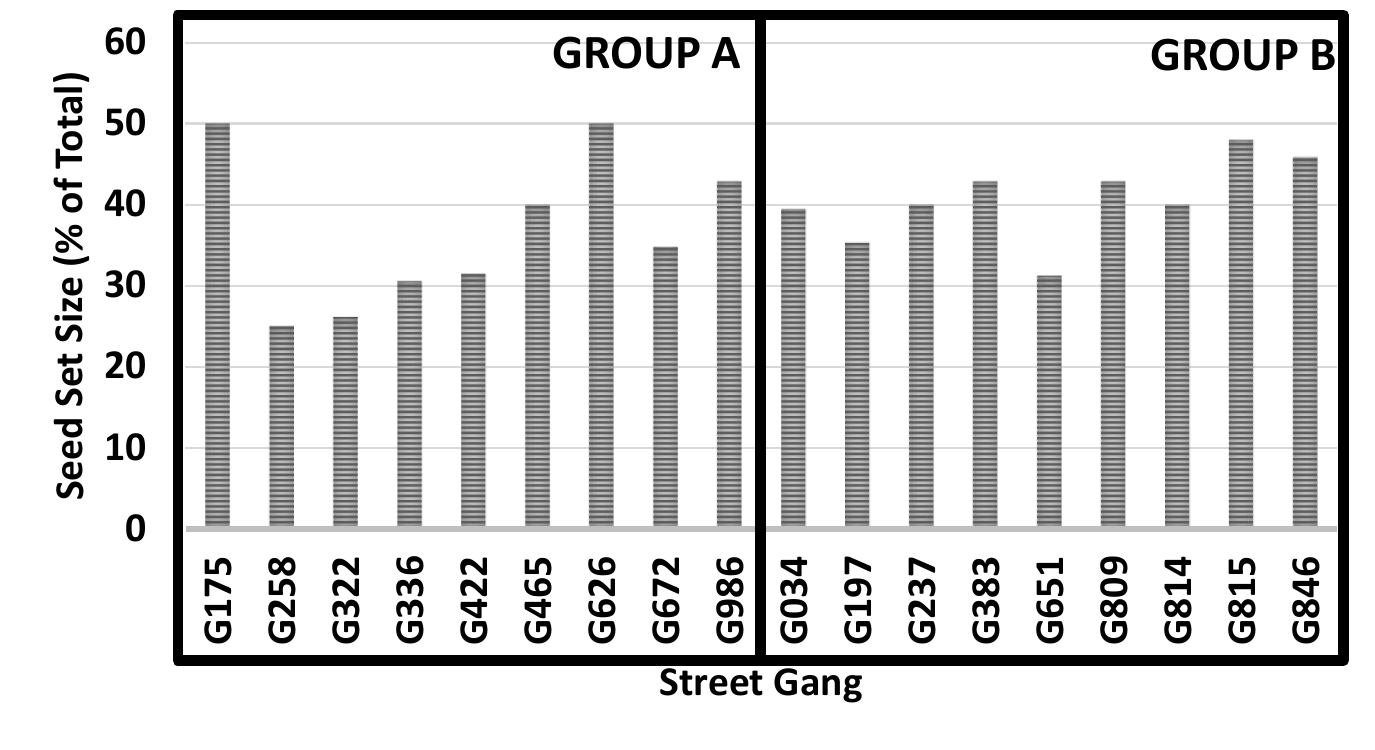}
\caption{Seed size as a percentage of the total gang membership for the $18$ street gangs analyzed by ORCA organized into two different racial groups.}
\label{seedSize}
\end{figure}

We also examined the seed sets found by ORCA for each of the street gangs.  These are sets of individuals who can initiate a social cascade (under the majority threshold tipping model) that will cause universal adoption in the gang.  The size of the seed set as a percentage of the population of each gang is shown in Figure~\ref{seedSize}.  The street gangs in the police district we examined were racially segregated - belonging to one of two races.  Police officers working in the district have told us that gangs of Racial Group A are known for a more centralized organizational structure while gangs of Racial Group B have adopted a decentralized model.  These groups are denoted in Figure~\ref{histoFig}.  We were able to quantify this observation as the gangs in Racial Group A had (on average) seed sets $3.86\%$ smaller than those in Group B.

\subsection{Finding Communities and Identifying Ecosystems}

\begin{figure}[t!]
\centering
\includegraphics[width=0.4\textwidth]{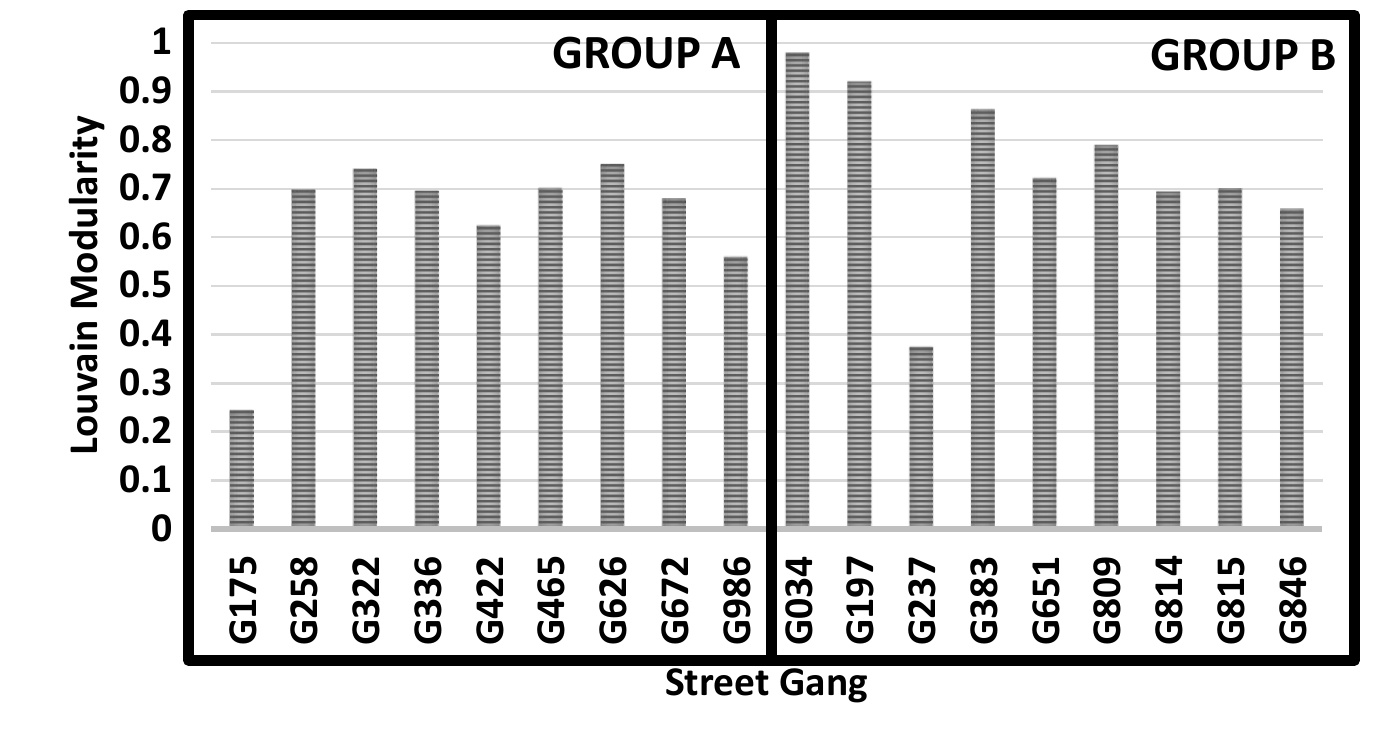}
\caption{The modularity of the partition found with the Louvain algorithm for each gang.}
\label{commStr}
\end{figure}

As stated earlier, law enforcement personnel have much interest in identifying sub-organizations of a given street gang.  ORCA tackled this problem using the Louvain algorithm as described in the previous section.  For the $18$ street gangs examined in this study, ORCA identified subgroups for each in an attempt to optimize modularity.  As modularity approaches $1.0$, the communities produced by the algorithm become more segregated.  We show the modularity values in Figure~\ref{commStr}.  Again, we also noticed a difference in the organizational structure based on the gang's race (aligning with anecdotal police observations).  Gangs associated with Racial Group A (more centralized) had lower modularity scores (by an average of $11.2\%$) than gangs associated with Racial Group B (decentralized).  This also corresponds with police observations that gangs affiliated with Racial Group B tend to operate in smaller factions (where the gang organizations server more as a confederation) as opposed to gangs affiliated with Group A (which are typically more hierarchical).

\begin{figure}[t!]
\centering
\includegraphics[width=0.55\textwidth]{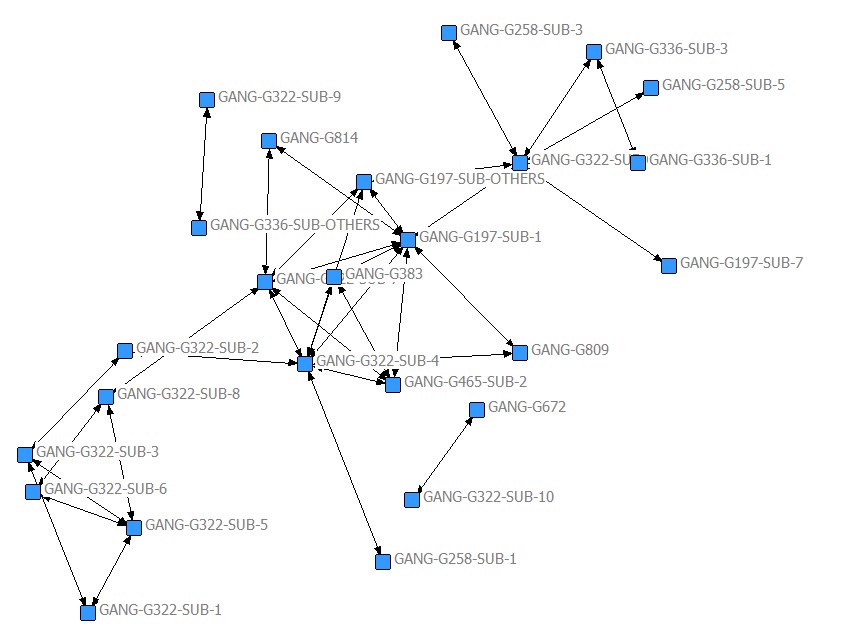}
\caption{An example ecosystem for one of the gangs analyzed by ORCA.}
\label{ecosys}
\end{figure}

\begin{figure}[t!]
\centering
\includegraphics[width=0.4\textwidth]{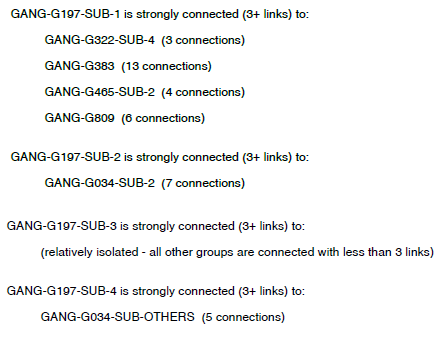}
\caption{An example ecosystem analysis for one of the gangs analyzed by ORCA.}
\label{bulletFig}
\end{figure}

\begin{figure}[t!]
\centering
\includegraphics[width=0.5\textwidth]{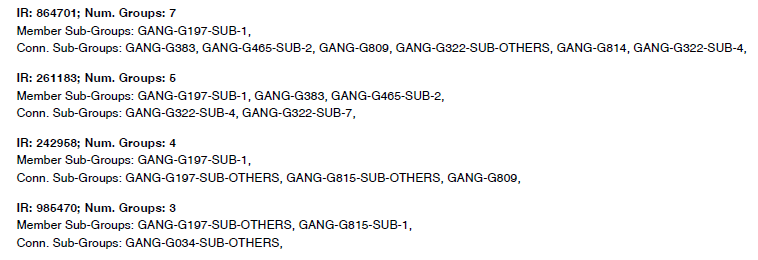}
\caption{An example report of connectors within one of the gangs analyzed by ORCA.}
\label{connFig}
\end{figure}

The creation of a gang's ecosystem (as described earlier) is derived directly from the result of the Louvain partition.  Sub-groups identified with the partition are connected together based on social links between members or by members who claim to be in more than one gang.  Such an ecosystem is shown in Figure~\ref{ecosys}.  ORCA also provides an analytical report of an ecosystem listing all sub-group relations and the strength of the relations (based on number of social ties between organizations) (see Figure~\ref{bulletFig}).  Additionally, ORCA also could identify individuals that connected various organizations.  Example output of this feature is shown in Figure~\ref{connFig}.

\subsection{User Interface}

\begin{figure}[t!]
\centering
\includegraphics[width=0.52\textwidth]{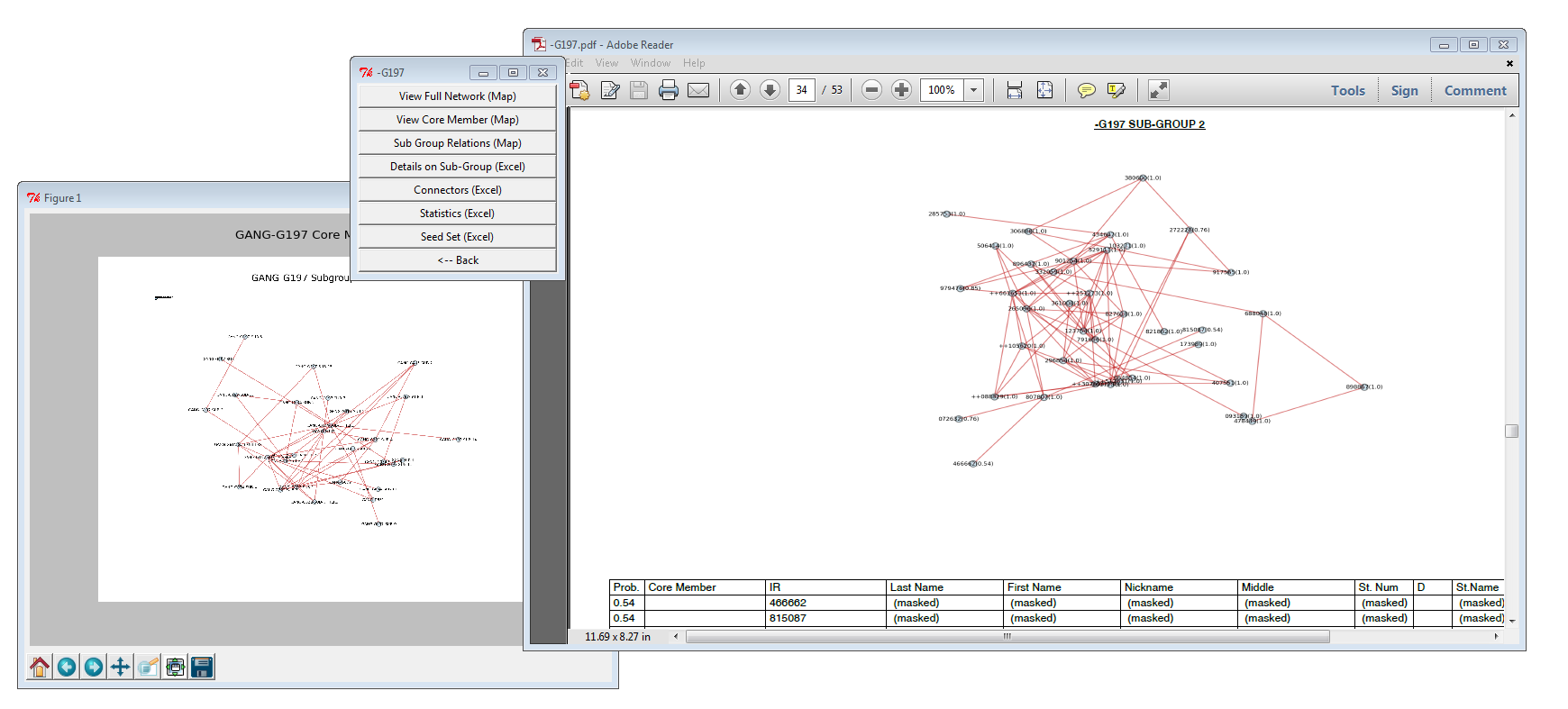}
\caption{The ORCA user interface with network visualization and PDF report output.}
\label{mainScreen}
\end{figure}

Another priority in developing ORCA was to make the software accessible to a variety of police personnel.  We created a user interface using the TKinter library version 8.5\footnote{http://wiki.python.org/moin/TkInter} and provided network visualizations using Matplotlib 1.2.0\footnote{http://matplotlib.org/}.  As displayed in Figure~\ref{mainScreen}, the users can utilize the interface to easily run new analysis using arrest and relationship data taken from the result of a database query.   They can also view the reports generated by completed analyisis.  When the analysis is complete, the law enforcement personnel have the ability to view a full report that is automatically generated in a portable document format (PDF) using version 1.7 of the PyFPDF library\footnote{https://code.google.com/p/pyfpdf/}.  This is useful because it is a widely used format that is familiar to most users.  In addition to the full report, the users can also view specific reports or network visualizations which allows them to easily view the results of the analysis that has been conducted.  Finally, we would like to note that this is a draft interface that we will refine throughout the summer of 2013 by conducting studies in the field as law enforcement personnel utilize the software.  We plan to incorporate the feedback that we gather into further versions of the software so that it becomes a more useful and usable tool.

\section{Related Work}
\label{rwSec}

There have been several other pieces of software introduced for both general purpose network analysis as well as for law enforcement.  The main contribution of the software presented in this paper if the ORCA software.  ORCA is that it is tailor made for specific police requirements dealing with violent street gangs.  Further, it leverage new articiial intellignece and data mining techniques such as MANCALog and \textsf{TIP\_DECOMP}.  Although ORCA is the first attempt to combine these techniques in a piece of intelligence analysis software, there have been other efforts to build software designed for social network-based intelligence analysis that we describe below.

Previous software designed to support law-enforcement through social network analysis include CrimeFighter~\cite{petersen11} and CrimeLink~\cite{sch11}.  However, these tools provide complementary capabilities to ORCA.  While ORCA is primarily designed for intelligence analysis with respect to street gangs, CrimeLink is designed to support investigations while CrimeFighter is directed more toward targeting of individuals in criminal organizations - rather than understanding membership in and linkages between sub-groups.  Further, neither of these software packages support degree of membership, seed set identification, or the creation of ecosystems.

General-purpose social network analysis software also exist, including the Organization Risk Analyzer (ORA)~\cite{Carley2004}, and NodeXL (an add-on to Microsoft Excel)~\cite{Hansen10}.  While these tools are very powerful and contain many features, they do not provide the degree of membership calculation or the identification of seed sets as ORCA does.  ORCA is also purposely designed for police intelligence analysts to better understand street gangs, and features such as community finding in ORCA are designed for with this application in mind (hence the creation of ecosystems and identification of inter-group connectors).  This nuanced use of community detection for street gang intelligence analysis is not ready for use ``out of the box'' in a general-purpose social network analysis package such as ORA or NodeXL.
\section{Conclusion}

In this paper we introduced ORCA - the Organizational, Relationship, and Contact Analyzer - a tool designed from the ground-up to support intelligence analysis to aide law enforcement personnel in combating violent street gangs.  We have shown how ORCA can meet various police analysis needs to include determining degree for gang members of unknown affiliation, identifying sets of influential individuals in a gang, and finding and analyzing the sub-organizations of a gang to determine inter/intra-group relationships.

Our next step with regard to this work is to integrate geospatial and temporal elements in the analysis - particularly with respect to community finding and degree of membership.  Currently we are working closely with a major metropolitan police department to transition ORCA for use in law enforcement.  Throughout the summer of 2013, we are sending project assistants to work closely with the police in order to identify additional police requirements that can be addressed with techniques similar to those discussed in this paper. Currently the police are employing this analysis for one district.  There are plans to expand to other districts in late 2013.

\section*{Acknowledgment}
The authors are supported under by the Army Research Office (project 2GDATXR042) and the Office of the Secretary of Defense.  The opinions in this paper are those of the authors and do not necessarily reflect the opinions of the funders, the U.S. Military Academy, or the U.S. Army.



\end{document}